%% file: main.tex
\documentclass[manuscript,screen,nonacm]{acmart} 
\acmConference[]{} 

\newcommand{\workshopname}{GenAICHI: CHI 2024 Workshop on Generative AI and HCI}
\newcommand{\licensedetails}{Licensed under a Creative Commons Attribution 4.0 International License (CC BY 4.0). Copyright remains with the author(s).}
\newcommand\extrafootertext[1]{
    \bgroup
    \renewcommand\thefootnote{\fnsymbol{footnote}}%
    \renewcommand\thempfootnote{\fnsymbol{mpfootnote}}%
    \footnotetext[0]{#1}%
    \egroup
}

\AtBeginDocument{ 
    \fancypagestyle{firstpagestyle}{
        \fancyhf{}
        \fancyfoot[L]{\sffamily\footnotesize \workshopname}%
        \fancyfoot[C]{\sffamily\footnotesize \thepage}
    }
    \fancyhf{}
    \fancyhead[L]{\sffamily\footnotesize\shorttitle}
    \fancyhead[R]{\sffamily\footnotesize\shortauthors}
    \fancyfoot[L]{\sffamily\footnotesize\workshopname}%
    \fancyfoot[C]{\sffamily\footnotesize\thepage}
    \extrafootertext{\licensedetails}
}

\begin{document}

\title[Generative AI in Color-Changing Systems]{Generative AI in Color-Changing Systems: Re-Programmable 3D Object Textures with Material and Design Constraints}

\author{Yunyi Zhu}
\email{yunyizhu@mit.edu}
\orcid{0000-0003-4545-8069}
\affiliation{%
  \institution{MIT CSAIL}
  \city{Cambridge}
  \country{USA}
}

\author{Faraz Faruqi}
\email{ffaruqi@mit.edu}
\orcid{0000-0002-1691-2093}
\affiliation{%
  \institution{MIT CSAIL}
  \city{Cambridge}
  \country{USA}
}

\author{Stefanie Mueller}
\email{stefanie.mueller@mit.edu}
\orcid{0000-0001-7743-7807}
\affiliation{%
  \institution{MIT CSAIL}
  \city{Cambridge}
  \country{USA}
}

\begin{abstract}
\input{sections/0_Abstract}

\end{abstract}


\newcommand{\yunyi}[1]{{\leavevmode\color[rgb]{0.0, 0.5, 1.0}{#1}}}



\maketitle


\input{sections/1_Introduction}
\input{sections/2_Photochromic}
\input{sections/3_Usability}
\input{sections/4_Conclusion}


\bibliographystyle{ACM-Reference-Format}
\bibliography{direction1, direction2}

\end{document}

%% file: sections/0_Abstract.tex
Advances in Generative AI tools have allowed designers to manipulate existing 3D models using text or image-based prompts, enabling creators to explore different design goals. Photochromic color-changing systems, on the other hand, allow for the reprogramming of surface texture of 3D models, enabling easy customization of physical objects and opening up the possibility of using object surfaces for data display. However, existing photochromic systems require the user to manually design the desired texture, inspect the simulation of the pattern on the object, and verify the efficacy of the generated pattern. These manual design, inspection, and verification steps prevent the user from efficiently exploring the design space of possible patterns. Thus, by designing an automated workflow desired for an end-to-end texture application process, we can allow rapid iteration on different practicable patterns. 

In this workshop paper, we discuss the possibilities of extending generative AI systems, with material and design constraints for reprogrammable surfaces with photochromic materials. By constraining generative AI systems to colors and materials possible to be physically realized with photochromic dyes, we can create tools that would allow users to explore different viable patterns, with text and image-based prompts. We identify two focus areas in this topic: photochromic material constraints and design constraints for data-encoded textures. We highlight the current limitations of using generative AI tools to create viable textures using photochromic material. Finally, we present possible approaches to augment generative AI methods to take into account the photochromic material constraints, allowing for the creation of viable photochromic textures rapidly and easily.

%% file: sections/1_Introduction.tex
\section{Introduction}

Generative AI tools have been used to create digital assets such as images~\cite{rombach2022high} and 3D models~\cite{get3d_gao_2022, jun2023shap}. For example, existing generative AI tools such as Text2Mesh~\cite{text2mesh_michel_2022} and Magic3D~\cite{magic3d_lin_2022} have enabled creating customized 3D models based on user's text or image prompts. These tools allow users to personalize or generate their creative designs with text or image prompts, achieving rapid and easy iteration for creating new designs. 

For 3D manipulation, these generative AI models are designed to align the visual quality of a model to that of the desired style input by the user. However, they do not consider fabrication constraints during geometry manipulation. As a result, while these tools can generate 3D models with the desired appearance in digital space, their functionality may not translate into the physical world. In digital fabrication, researchers have proposed Style2Fab~\cite{style2fab_faruqi_2023}, which allows users to retain their desired functionality in 3D models while personalizing them based on their desired style~\cite{style2fab_faruqi_2023}. This tool takes into account the post-fabrication functional constraints of the model while modifying the 3D model with generative AI, enabling the customization of objects that are functional in the physical world. 

Nevertheless, 3D printing as a fabrication method has its limitations, specifically, the long wait time between outputs. Re-fabricating an object for customizing an existing object is a time and material-consuming process. In contrast, using color-changing material allows quick manipulation of the surface texture of an object, enabling much faster visual customization and iteration on existing objects. For instance, Photo-Chromeleon~\cite{photochromeleon}, has allowed reprogramming the surface texture of an object within an hour using a projection system. Using photochromic material, whose color can be reprogrammed using a projection system, Photo-Chromeleon supports users in customizing 3D physical objects with their desired patterns. 



While color-changing materials, such as photochromic material, allow for faster and more sustainable customization on physical objects, they also bring a unique set of constraints. In this workshop paper, we explore ways to augment generative AI algorithms to address challenges in photochromic color-changing systems to enable automated re-programmable textures on surfaces of 3D objects. We discuss the material constraints from the photochromic material and design constraints from data-encoded textures. In particular, we discuss how to generate textures that optimize for the photochromic color space, color application time, and light source form factor and are visible, legible, and data-driven. 



%% file: sections/2_Photochromic.tex
\section{Material Constraints in Color-Changing Systems}

Since color-changing systems depend on physical properties of the material to achieve each texture, the range of achievable textures is subject to specific restrictions inherent to each material's properties. In the example of photochromic material as in Photo-Chromeleon~\cite{photochromeleon}, the achievable texture has a more constrained color space compared to the RGB color space commonly used in image generation. Different textures can have different color-changing time, adding another factor to the decision process. The designed light source can post additional restrictions on the color space, resolution, and object geometry.  

Below, we discuss each of the constraints in detail.  

\vspace{0.5em}
\noindent
\textbf{Available Color Space: } The color space of the photochromic color-changing systems, i.e. the color that can be achieved by reprogramming the photochromic material, is smaller than the traditional RGB color space. This is because the three color channels are partially dependent, i.e. the same wavelength of light can react with multiple color channels at the same time. But generative AI models, specifically text-to-image models~\cite{rombach2022high} utilize the entire RGB space to maximize the visual quality of the output. By constraining these text-to-image models to the color space achievable by photochromic material, we can maximize the visual quality of the photochromic result. Therefore, when generating texture for photochromic material, the generative AI algorithm should keep the design within the available color space to avoid generating designs that cannot be realized on the physical object. 

\vspace{0.5em}
\noindent
\textbf{Color Application Time: } In photochromic color-changing systems, the time to apply the design depends on the design itself. This is because each of the cyan, magenta, and yellow color channels requires a different amount of time to achieve the target saturation with its corresponding light wavelength. Thus, some patterns, depending on the colors, would take significantly more or less time to finish the reprogramming process. By augmenting a generative AI tool to take a time factor along with a design prompt, we can enable to generation of time-sensitive patterns. Such a generative tool can be trained by a synthetic dataset of patterns and their corresponding saturation times, which can in turn be used to generate time-efficient patterns. 


\vspace{0.5em}
\noindent
\textbf{Type of Light Source: } Different types of light sources, including ones based on projectors~\cite{photochromeleon, chromoupdate} and LEDs~\cite{chromocloth}, have been used in multi-color photochromic color-changing systems. In existing systems, projection-based light sources have higher resolution and doubly-curved surfaces, while LED-based light sources have shorter color-changing times and are more portable. Therefore, when generating the texture, the algorithm should take into account the light source and generate a texture with the corresponding resolution, speed requirement, and geometry constraints according to the light source.




%% file: sections/3_Usability.tex
\section{Data-Encoded Re-Programmable Texture Generation}
Color-changing systems open up the possibility to encode information on the surface of the object. Due to the short time needed to apply a new texture, the surface of a photochromic object can not only function for aesthetic purposes but also serve as a dynamic display for information-rich, data-encoded textures.

Existing photochromic color-changing systems require significant manual input from users to achieve the desired texture on a specific object\cite{photochromeleon, chromoupdate}. In particular, the user needs to manually design the 3D model's texture as a PNG image, put the text and charts on the correct location of the image, and use UV mapping to map the texture onto an object. This manual design process can be time-consuming and interrupts the seamless cycle of updating surface textures. 

However, the identification and generative power of recent generative AI models present a unique opportunity for creating a seamless design process for photochromic-based surface texture updates. While there exist generative AI tools that achieve texture generation for an existing model, they focus on stylization instead of data presentation. Below, we point out several key challenges in generating data-encoded textures for photochromic color-changing objects: 

\vspace{0.5em}
\noindent
\textbf{Identifying Viable Regions for Presenting Information:} To generate data-encoded texture on the surface of a 3D object, it is important to place the information in visible regions of the object. For example, the bottom of a mug and the inside of a jar should not display any information as the user will not be able to see it. Generative AI approaches to 3D content creation are efficient at reconstructing 3D models from scratch, based on a text or image prompt. Thus, to achieve visible presentation, a generative AI model can be leveraged to find viable sections of the input model based on the data or pattern to be presented, and only generate text-encoded information on those sections.

\vspace{0.5em}
\noindent
\textbf{Adjusting Visualization Size and Orientation:} Different from standard style generation, data-driven visualization such as text display, needs to be rendered in a legible size and orientation when transferred to the physical object. For example, the same text should be of different sizes and positions when generated on the surface of a mug versus on a wall. To achieve this, the generative AI system needs to have an accurate representation and the position of the input 3D model, including information on its size and distance from the projection system, and how are textures designed for such objects.

\vspace{0.5em}
\noindent
\textbf{Generating Texture Style based on User Data:} In an end-to-end color-changing system powered by generative AI, the texture can be generated passively from data collected from the environment. This is different from existing generative AI tools that generate styles based on users' intentional text input. To achieve this, the system can incorporate data visualization~\cite{2023-vistext} algorithms that act as input data and create a texture style that forms a visualization from the input data.

%% file: sections/4_Conclusion.tex
\section{Conclusion}
The goal of this position paper is to discuss how generative AI can be used in conjunction with photochromic color-changing systems to create an end-to-end color-changing system that allows for seamless environment reprogramming and data embedding into an ambient environment. We discussed potential constraints specific to photochromic color-changing interfaces and data-encoded object texture generation. We also discussed how to augment existing generative AI models to yield textures that optimize for the color space, application time, and light source form factor and are visible, legible, and data-driven.